\documentclass[twocolumn,showpacs,preprintnumbers]{revtex4}
\usepackage{graphicx}
\usepackage{dcolumn}
\usepackage{bm}
\begin{document}
\newcommand{\ds}{\displaystyle}
\newcommand{\be}{\begin{equation}}
\newcommand{\en}{\end{equation}}
\newcommand{\bea}{\begin{eqnarray}}
\newcommand{\ena}{\end{eqnarray}}
\title{N-dimensional static and evolving Lorentzian wormholes with cosmological constant}
\author{Mauricio Cataldo}
\altaffiliation{mcataldo@ubiobio.cl} \affiliation{Departamento de
F\'\i sica, Facultad de Ciencias, Universidad del B\'\i o--B\'\i
o, Avenida Collao 1202, Casilla 5-C, Concepci\'on, Chile.}
\author{Paola Meza}
\altaffiliation{paolameza@udec.cl} \affiliation{Departamento de
F\'\i sica, Facultad de Ciencias F\'\i sicas y Matem\'aticas,
Universidad de Concepci\'on, Casilla 160-C, Concepci\'on, Chile.}
\author{Paul Minning}
\altaffiliation{pminning@udec.cl} \affiliation{Departamento de F\'\i
sica, Facultad de Ciencias F\'\i sicas y Matem\'aticas, Universidad
de Concepci\'on, Casilla 160-C, Concepci\'on, Chile.}
\date{\today}
\begin{abstract}
{\bf {Abstract:}} We present a family of static and evolving
spherically symmetric Lorentzian wormhole solutions in N+1
dimensional Einstein gravity. In general, for static wormholes, we
require that at least the radial pressure has a barotropic equation
of state of the form $p_r=\omega_r \rho$, where the state parameter
$\omega_r$ is constant. On the other hand, it is shown that in any
dimension $N \geq 3$, with $\phi(r)=\Lambda=0$ and anisotropic
barotropic pressure with constant state parameters, static wormhole
configurations are always asymptotically flat spacetimes, while in
2+1 gravity there are not only asymptotically flat static wormholes
and also more general ones. In this case, the matter sustaining the
three-dimensional wormhole may be only a pressureless fluid. In the
case of evolving wormholes with $N \geq 3$, the presence of a
cosmological constant leads to an expansion or contraction of the
wormhole configurations: for positive cosmological constant we have
wormholes which expand forever and, for negative cosmological
constant we have wormholes which expand to a maximum value and then
recollapse. In the absence of a cosmological constant the wormhole
expands with constant velocity, i.e without acceleration or
deceleration. In 2+1 dimensions the expanding wormholes always have
an isotropic and homogeneous pressure, depending only on the time
coordinate.

\vspace{0.5cm} 
\end{abstract}
\smallskip\
\maketitle \preprint{}
\section{Introduction}
It is well known the interest of studying gravitational fields in
spacetimes with arbitrary dimensions, with or without cosmological
constant. The theoretical properties of these multidimensional
gravitational fields could be quite different from one dimension to
another and so it is of much interest to get an insight into how the
space-time dimension may influence the gravitational dynamics. For
example, recently, there was a lot of interest in low-dimensional
gravity. The discovery of the existence of three-dimensional black
hole solutions represents one of the main advances for
low-dimensional gravity theories. While in (3+1)-dimensional gravity
black hole solutions there exist with or without cosmological
constant, in (2+1) dimensions the cosmological constant plays a clue
role in their existence. From the point of view of the equilibrium
configurations of stars the number of dimensions of space–time also
can influence this equilibrium. In Ref.~\cite{Norman} the authors
show that dimensionality does increase the effect of mass but not
the contribution of the pressure, which is the same in any
dimension.

The efforts also have been directed to the extension of the analysis
of (3+1)-solutions to higher dimensional spacetimes, which also have
attracted the attention of the community. The interest is related
for example to black hole physics~\cite{Gallo}, wormhole physics,
Kaluza-Klein gravity, multidimensional and string/brane cosmology,
among others. It is interesting to note for example, that there is a
lack of uniqueness for black holes in higher dimensions, unlike the
four dimensional counterparts. Specifically, the higher dimensional
rotating black hole metric~\cite{Myers} is not unique, unlike the
Kerr geometry in (3+1) dimensions~\cite{Emparan}.

The showed interest in higher dimensional spacetimes can be extended
also to the study of wormhole physics, which has rapidly grown into
an active area of research. Euclidean wormholes have been studied by
Gonzales–Diaz and by Jianjun and Sicong~\cite{Gonzales–Diaz} for
example. The Lorentzian ones have been studied in the context of the
N-dimensional Einstein gravity~\cite{Cataldo3} and
Einstein-Gauss-Bonnet theory of gravitation~\cite{Bhawal}. Wormholes
in the context of brane worlds are discussed in~\cite{Rodrigo},
while the construction of thin-shell electrically charged wormholes
in d-dimensional general relativity with a cosmological constant is
discussed in Ref.~\cite{Lemos}.

Evolving higher dimensional wormholes have been studied in
Refs.~\cite{DeBenedictis} and~\cite{Kar}. The authors of
Ref.~\cite{DeBenedictis} study wormhole solutions to Einstein
gravity with an arbitrary number of time dependent compact
dimensions and a matter-vacuum boundary. On the other hand, in
Ref.~\cite{Kar} the authors consider non-static wormholes, mainly in
2+1 and 3+1 dimensions, with the required matter satisfying the weak
energy conditions. The authors explore several different scale
factors and derive the corresponding consequences.

In this paper, we shall obtain a family of static and evolving
spherically symmetric wormhole solutions, in N+1 dimensional
gravity, in the presence of a cosmological constant and by imposing
at least a barotropic equation of state, with constant state
parameter, on the radial pressure.

The organization of the paper is as follows: In Sec. II we give some
characterization of Lorentzian wormholes. In Sec. III the field
equations for evolving wormholes in N+1 gravity are formulated. In
Sec. IV and Sec. V the static and non-static N+1 dimensional
wormholes are discussed respectively for $N \geq 3$. In Sec. VI the
2+1-dimensional static and non-static wormholes are treated.
Finally, in Sec. VII we conclude with some remarks.

\section{Characterization of Lorentzian wormholes}
On the purely gravitational side, the interest on wormhole
geometries has been mainly focused on Lorentzian wormholes, and was
especially stimulated by the pioneering work of Morris, Thorne and
Yurtsever~\cite{Morris1}, where static, spherically symmetric
Lorentzian wormholes were defined and considered to be an exciting
possibility for constructing time machine models with these exotic
objects. The metric ansatz of Morris and Thorne~\cite{Morris} for
the spacetime which describes a static Lorentzian wormhole is given
by
\begin{eqnarray}\label{4wormhole}
ds^2=-e^{2\phi(r)}dt^2+\frac{dr^2}{1-\frac{b(r)}{r}}+r^2 d
\Omega_2^2,
\end{eqnarray}
where $d \Omega_2^2=d\theta^2+sin^2 \theta d \varphi^2$ and the
functions $\phi(r)$ and $b(r)$ are referred to as redshift function
and shape function respectively.

Morris and Thorne have discussed in detail the general constraints
on the functions $b(r)$ and $\phi(r)$ which make a
wormhole~\cite{Morris,Visser}:

Constraint 1: A no--horizon condition, i.e. $e^{\phi(r)}$ is
finite throughout the space--time in order to ensure the absence
of horizons and singularities.

Constraint 2: Minimum value of the r--coordinate, i.e. at the
throat of the wormhole $r=b(r)=b_0$, $b_0$ being the minimum value
of $r$.

Constraint 3: Finiteness of the proper radial distance, i.e.
\begin{eqnarray}\label{r finito}
\frac{b(r)}{r} \leq 1,
\end{eqnarray}
(for $r \geq b_0$) throughout the space--time. This is required in
order to ensure the finiteness of the proper radial distance
$l(r)$ defined by
\begin{eqnarray}
l(r)=\pm \int^r_{b_0} \frac{dr}{\sqrt{1-b(r)/r}}.
\end{eqnarray}
The $\pm$ signs refer to the two asymptotically flat regions which
are connected by the wormhole. The equality sign in~(\ref{r
finito}) holds only at the throat.

Constraint 4: Asymptotic flatness condition, i.e. as $l \rightarrow
\pm \infty$ (or equivalently, $r \rightarrow \infty$) then $b(r)/r
\rightarrow 0$ and $\phi(r) \rightarrow 0$.

Notice that these constraints provide a minimum set of conditions
which lead, through an analysis of the embedding of the spacelike
slice of~(\ref{4wormhole}) in a Euclidean space, to a geometry
featuring two asymptotically flat regions connected by a
bridge~\cite{Dadhich}.

In this paper we are not including considerations about the
traversability constraints discussed by Morris and
Thorne~\cite{Morris}.

\section{N+1--dimensional Lorentzian wormholes}

\subsection{The metric and the matter source}

The evolving spherically symmetric wormhole in higher dimensions
may be obtained by a simple generalization of the original Morris
and Thorne metric~\cite{Morris} to a time-dependent metric given by
\begin{eqnarray}
ds^2=-e^{2\phi(t,r)}dt^2+a^2(t)\left(\frac{dr^2}{1-\frac{b(r)}{r}}+
r^2 d\Omega^2_{N-1}\right) = \nonumber \\
-\theta^{(t)}\theta^{(t)}+\theta^{(r)}\theta^{(r)}+\sum_{i=1}^{N-1}
\theta^{(\theta_i)}\theta^{(\theta_i)}.
\,\,\,\,\,\,\,\,\,\,\,\,\,\,\, \label{1}
\end{eqnarray}
where $a(t)$ is the scale factor of the universe, and
$\theta^{(\mu)}$  are the proper orthonormal basis whose one-forms
are given by,
\begin{eqnarray*}
 \theta^{(t)}=e^{\phi(t,r)} \, dt,
\end{eqnarray*}
\begin{eqnarray*}
 \theta^{(r)}=a(t)\frac{dr}{\sqrt{1-\frac{b(r)}{r}}},
\end{eqnarray*}
\begin{eqnarray*}
 \theta^{(\theta_1)}=a(t)r \,d\theta_1,
\end{eqnarray*}
\begin{eqnarray*}
 \theta^{(\theta_2)}=a(t)r \sin\theta_1 \,d\theta_2,\ldots,
\end{eqnarray*}
\begin{eqnarray*}
 \theta^{(\theta_{N-1})}=a(t)r \,\prod_{i=1}^{N-2}\sin\theta_i \,d\theta_{N-1}.
\end{eqnarray*}
It must be noticed that the metric~(\ref{1}) includes static
wormholes determined by the condition $a(t)=a_0=const$. The constant
$a_0$ may be absorbed by redefining the shape function in the
following form: $b(r)\mapsto r-a_0^2 (r-b(r))$.

In general, for the metric~(\ref{1}), one might introduce a matter
source described by an imperfect fluid. Of course in this case the
energy-momentum tensor has also non-diagonal entries. However, we
shall use the notion of phantom energy in a slightly more extended
sense: We shall consider this traditionally homogeneous and
isotropic exotic source to be generalized to an inhomogeneous and
anisotropic fluid, but still with a diagonal energy-momentum tensor.
This means that the only nonzero components of the energy-momentum
tensor in this basis are
\begin{equation}\label{T}
T_{(t)(t)}=\rho(t,r), \, T_{(r)(r)}=p_r(t,r)=-\tau(t,r),
\end{equation}
and
\begin{equation}\label{TT}
 T_{(\theta_1)(\theta_1)}=\ldots=T_{(\theta_{N-1})(\theta_{N-1})}=p_l(t,r).
\end{equation}
where the quantities $\rho(t,r)$, $p_r(t,r)$,
$\tau(t,r)(=-p_r(t,r))$, and $p_l(t,r)(=p_{\theta_i}(t,r)$) are
respectively the energy density, the radial pressure, the radial
tension per unit area, and the lateral pressure as measured by
observers who always remain at rest at constant $r$, $\theta_i$.

\subsection{The Einstein field and the conservation equations}
For the evolving spherically symmetric wormhole metric (\ref{1}) the
Einstein field equations with cosmological constant $\Lambda$ are
given by
\begin{eqnarray}
\frac{N\left(N-1\right)}{2} e^{-2\phi(t,r)} H^2 &+& \nonumber \\
+\frac{\left(N-1\right)}{2a^2
r^3}\left(rb'+\left(N-3\right)b\right)&=& \kappa \rho(t,r) +
\Lambda, \label{9}
\end{eqnarray}
\begin{equation}\label{2}
\frac{\left(N-1\right)e^{-\phi(t,r)}H}{a}\sqrt{\frac{r-b(r)}{r}}\frac{\partial
\phi}{\partial r}=0,
\end{equation}
\begin{eqnarray}
-\left(N-1\right)e^{-2\phi(t,r)}\left(\frac{N-2}{2}H^2 + \frac{\ddot{a}}{a}-
H\frac{\partial \phi}{\partial t}\right)&-& \nonumber \\
-\frac{\left(N-1\right)\left(N-2\right)}{2r^3 a^2}b +
\frac{\left(N-1\right)\left(r-b(r)\right)}{r^2 a^2}\frac{\partial \phi}{\partial r}&=& \nonumber \\
=\kappa p_r(t,r)-\Lambda,&& \label{3}
\end{eqnarray}
\begin{eqnarray}
-\left(N-1\right)e^{-2\phi(t,r)}\left(\frac{N-2}{2}H^2+\frac{\ddot{a}}{a}-
H\frac{\partial \phi}{\partial t}\right)&-& \nonumber \\
-\frac{rb'+\left(2N-5\right)b-\left(2N-4\right)r}{2a^2 r^2}\frac{\partial \phi}{\partial r}&-& \nonumber \\
-\frac{N-2}{2r^3a^2}\left(rb'+\left(N-4\right)b\right)&+& \nonumber \\
+\frac{\left(r-b\right)}{a^2 r}\left(\frac{\partial^2 \phi}{\partial
r^2}+\left(\frac{\partial \phi}{\partial r}\right)^2\right)=\kappa
p_l(t,r)-\Lambda,  \label{4}
\end{eqnarray}
where $\kappa=8\pi G$, $H=\dot{a}/a$ and an overdot and a prime
denote differentiation $d/dt$ and $d/dr$ respectively. Using the
conservation equation $\nabla_{\mu}T^{\mu}_{\phantom{1}\nu}=0$ we
have that,
\begin{eqnarray}
\frac{\partial \rho}{\partial t}+H\left(p_r + (N-1)p_l + N\rho\right)&=&0, \label{ConsEQ1}\\
\frac{\partial p_r}{\partial r}+\frac{\partial \phi}{\partial
r}\left(\rho + p_r\right)-\frac{N-1}{r}\left(p_l-p_r\right)&=&0.
\label{ConsEQ2}
\end{eqnarray}
\paragraph*{}
We shall study matter sources described with at least a barotropic
equation of state for $p_r(t,r)$. Thus we can write for the radial
pressure
\begin{equation}
p_r(t,r)=\omega_r \rho(t,r), \label{12}
\end{equation}
where $\omega_r$ is a constant state parameter. One can require the
same for $p_l(t,r)$. Thus in some cases we shall consider solutions
with a lateral pressure given by
\begin{equation}
p_l(t,r)=\omega_l \rho(t,r), \label{12A}
\end{equation}
where $\omega_l$ is a constant state parameter.

In order to find solutions to the field equations we can see that
the equation~(\ref{2}) plays a fundamental role. For the diagonal
energy-momentum tensor~(\ref{T}) and~(\ref{TT}), Eq.~(\ref{2})
implies that the solutions are separated into two branches: one
static branch given by the condition $H=\dot{a}/a=0$ and another
non-static branch for $\partial \phi(t,r)/\partial r=0$.

\section{Static $N+1$ Wormhole solutions}
In general, for the static case, we shall suppose that the shift and
the shape functions, energy density and pressures are functions of
the radial coordinate $r$, and that only the radial pressure has a
barotropic equation of state given by $p_r(r)=\omega_r \rho(r)$.
Then, the required condition for the static branch $H=\dot{a}/a=0$
implies that the field equations take the following form:
\begin{eqnarray}
\frac{\left(N-1\right)}{2r^3}\left(rb'+\left(N-3\right)b\right)&=&
\kappa \rho(r) + \Lambda, \label{static9}
\end{eqnarray}
\begin{eqnarray}
-\frac{\left(N-1\right)\left(N-2\right)b}{2r^3} +\frac{\left(N-1\right)\left(r-
b(r)\right)}{r^2}\frac{d \phi}{d r} \nonumber \\
=\kappa \omega_r \rho(r)-\Lambda, \label{static3}
\end{eqnarray}
\begin{eqnarray}
-\frac{rb'+\left(2N-5\right)b-\left(2N-4\right)r}{2r^2}\frac{d
\phi}{d r} -\frac{N-2}{2r^3} \times \nonumber \\ \left(rb'+\left(N-4\right)b\right) 
+\frac{\left(r-b\right)}{r}\left(\frac{d^2 \phi}{d
r^2}+\left(\frac{d \phi}{d r}\right)^2\right)
\\ \nonumber =\kappa p_l(r)-\Lambda.   \label{static4}
\end{eqnarray}
In this case the conservation equation takes the form
\begin{eqnarray}\label{staticconseq}
\omega_r \frac{d \rho}{d r}+\frac{d \phi}{d r}\left(1 + \omega_r
\right) \rho-\frac{N-1}{r}\left(p_l-\omega_r \rho \right)=0.
\end{eqnarray}
In order to solve the field equations we can consider only
Eqs.~(\ref{static9}),(\ref{static3}) and~(\ref{staticconseq}). Thus
we have four unknown functions of r, i.e. $\rho(r)$, $p_l(r)$,
$b(r)$ and $\phi(r)$, for three field equations. In order to
construct solutions one can consider restricted choices for $b(r)$
or $\phi(r)$. So, we shall construct solutions by finding the
lateral pressure $p_l(r)$ from the conservation
equation~(\ref{staticconseq}), the energy density $\rho(r)$ from
Eq.~(\ref{static9}), and the shape function $b(r)$ (giving a
restricted form of $\phi(r)$) or the redshift function $\phi(r)$
(giving a restricted form of $b(r)$) from a differential equation
obtained from Eqs.~(\ref{static9}) and~(\ref{static3}). Thus, from
Eqs.~(\ref{static9}) and~(\ref{static3}) we find that the shape
function may be written in the form
\begin{widetext}
\begin{eqnarray}\label{eeb}
b(r)= \frac{Cr^{\frac{2-N}{\omega_r}-(N-3)}}{e^{2\phi(r)/\omega_r}}
+ \frac{2r^{\frac{2-N}{\omega_r}-(N-3)}}{\omega_r (N-1)
e^{2\phi(r)/\omega_r}} \int \left(\frac{}{}
(N-1)\phi^{\prime}+\Lambda (1+\omega_r) r \right)
r^{\frac{(N-2)(\omega_r+1)}{\omega_r}} e^{\frac{2\phi(r)}{\omega_r}}
dr,
\end{eqnarray}
or equivalently the redshift function as
\begin{eqnarray}\label{eephi}
\phi(r)=\int \frac{\left( \omega_{{r}} \left( N-3 \right) +N-2
 \right)\left( N-1 \right) b \left( r \right) -2\, \left( 1+\omega_{{r}} \right) \Lambda
\,{r}^{3}+ \left( N-1 \right) \omega_{{r}}r \,
b{\prime}}{2(N-1)(r-b(r))r}.
\end{eqnarray}
\end{widetext}

Now we shall consider specific static wormhole solutions.

\subsection{$b(r) \sim r^{\alpha}$ solution}
First, let us consider the case $b(r)=r_0 \, (r/r_0)^{\alpha}$ with
$\Lambda=0$. This choice permits us to consider the possibility of
having an asymptotically flat space-time. By putting these
expressions into Eq.~(\ref{eephi}) we have that the shift function
takes the form
\begin{eqnarray}\label{metric230}
e^{\phi(r)}=\left( 1-{\left(\frac{r}{r_0}\right)}^{-1+\alpha}
\right) ^{{\frac {N \left( 1+\omega_{{r}} \right)
-2-3\,\omega_{{r}}+\omega_{{r}}\alpha}{2(1-\alpha)}}},
\end{eqnarray}
then the metric and the energy density are given by
\begin{eqnarray}\label{metric23}
ds^2=-\left( 1-{\left(\frac{r_0}{r}\right)}^{1-\alpha} \right)
^{{\frac {N \left( 1+\omega_{{r}} \right)
-2-3\,\omega_{{r}}+\omega_{{r}}\alpha}{1-\alpha}}}dt^2 + \nonumber
\\ \frac{dr^2}{1-\left(\frac{r_0}{r}\right)^{1-\alpha}}+r^2 d \Omega_{N-1}^2,
\end{eqnarray}
\begin{eqnarray}\label{rho230}
\kappa\,\rho \left( r \right) =\frac{\left( N-1 \right) \left(
\alpha +N-3 \right)}{2 {r_{{0}}}^{2} } \left( {\frac {r_0}{r}}
\right) ^{3-\alpha},
\end{eqnarray}
respectively. Note that by making $N=3$ we obtain the 3+1-wormhole
solution discussed by Lobo in Ref.~\cite{Lobo}.

From the metric~(\ref{metric23}) we conclude that we have an
asymptotically flat space-time if $1-\alpha>0$ and ${{{N
\left(1+\omega_{{r}} \right) -2-3\,\omega_{{r}}+\omega_{{r}}\alpha}
}}>0$, since in this case we have that $e^{\phi(r)} \rightarrow 1$
and $b(r)/r \rightarrow 0$ for $r \rightarrow \infty$. However this
N-dimensional space-time is a non-traversable wormhole since an
event horizon is located at $r = r_0$. It is interesting to note
that for $N \geq 3$ this non-traversable wormhole may have a
positive energy density by requiring that $-(N-3)<\alpha<1$. In
order to make this space-time a traversable wormhole we must require
that $e^{\phi(r)}=1$ which implies that ${{{N \left(1+\omega_{{r}}
\right) -2-3\,\omega_{{r}}+\omega_{{r}}\alpha} }}=0$, or
equivalently $\alpha=\frac{2+3\,\omega_{{r}}-N
(1+\omega_{{r}})}{\omega_r}$. Finally, the metric and the energy
density take the following form:
\begin{eqnarray}\label{metric015}
ds^2=dt^2 - \frac{dr^2}{1-
\left(\frac{r_0}{r}\right)^{\frac{(1+\omega_r)(N-2)}{\omega_r}}}-
r^2 d \Omega_{N-1}^2,
\end{eqnarray}
\begin{eqnarray}\label{rho015}
\kappa\,\rho \left( r \right) =\frac{\left( N-1 \right) \left( 2-N
\right)}{2 \omega_r {r_{{0}}}^{2} } \left( {\frac {r_0}{r}} \right)
^{\frac{N(1+\omega_r)-2}{\omega_r}}.
\end{eqnarray}
Among $p_r(r)=\omega \rho(r)$, we have for the lateral pressure
\begin{eqnarray}\label{pl15}
\kappa p_{{l}}(r) =\frac{\left( N-2 \right)  \left( \omega_{{r
}}+N-2 \right)}{2 {\omega_{{r}}} r_0^{2}} {\left(\frac{r_0}{r}
\right)}^{{\frac {N \left( \omega_{{r}}+1 \right) -2}{
\omega_{{r}}}}}.
\end{eqnarray}
Note that for $\omega_r<-1$ and $\omega_r>0$ we have asymptotically
flat wormholes, while for $-1 < \omega_r < 0$ this does not occur.
On the other hand, for dimensions $N \geq 3$ we have a positive
energy density if $\omega_r <0$, and a negative energy density for
$\omega_r>0$. Thus we conclude that for $\omega_r>0$ we have
asymptotically flat wormholes with negative energy density, while
for $\omega_r<-1$ (or $-1<\omega_r<0$) asymptotically flat wormholes
supported by an energy of phantom (or quintessence) type, since we
have a negative radial pressure $p_r$. In this case, for $N \geq 3$,
the energy density $\rho \rightarrow 0$ for $r \rightarrow \infty$
since for $\omega_r<-1$ ($\omega_r>0$) always the inequality
$N(1+\omega_r)-2<0$ ($N(1+\omega_r)-2>0$) takes place.

Non asymptotically flat solutions may be obtained by requiring
$\Lambda \neq 0$. These spacetimes are expressed through
hypergeometric functions.

\subsection{$\phi(r)=0$ solution}
Now we shall consider the general static solution for a barotropic
radial pressure of the form~(\ref{12}) with $\phi(r)=0$. Clearly,
this is the more natural choice for the shift function, in order to
have a finite $e^{\phi(r)}$ throughout all the space--time. Thus by
putting $\phi(r)=0$ into Eq.~(\ref{eeb}) we find that
\begin{eqnarray}
b(r)=r_0\left(\frac{r}{r_0}
\right)^{\frac{2-N}{\omega_r}-(N-3)}+{\frac {2\Lambda \left(
\omega_{{r}}+1 \right)\,{r}^{3} }{ \left( \omega_{{r}}N+N-2 \right)
\left( N-1 \right) }}, \nonumber \\
\end{eqnarray}
and from Eqs.~(\ref{static3}) and~(\ref{staticconseq}) we obtain for
the energy density and the lateral pressure the following
expressions:
\begin{eqnarray}
\kappa \rho( r)=\frac{\left( N-1 \right)  \left(2- N \right)}{2
\omega_r r_0^{2}} \,{\left(\frac{r_0}{r}\right)}^{{\frac
{N(\omega_{{r}}+1)-2}{\omega_{{r}}}}} + \nonumber \\
{\frac {2\Lambda}{ \left( N(\omega_{{r}}+1)-2 \right) }},
\end{eqnarray}
\begin{eqnarray}
\kappa p_{{l}}(r) =\frac{\left( N-2 \right)  \left( \omega_{{r
}}+N-2 \right)}{2 {\omega_{{r}}} r_0^{2}} {\left(\frac{r_0}{r}
\right)}^{{\frac {N \left( \omega_{{r}}+1 \right) -2}{
\omega_{{r}}}}} +\nonumber
\\ {\frac {2\Lambda\, \omega_{{r}}}{\left( N \left(
\omega_{{r}}+1 \right) -2 \right) }}.
\end{eqnarray}
Clearly for $\Lambda=0$ we obtain the previous asymptotically flat
wormhole solution given by Eqs.~(\ref{metric015})-(\ref{pl15}). For
$\Lambda \neq 0$ we do not have two asymptotically flat regions, and
may have wormholes with two asymptotically de Sitter regions or two
asymptotically anti-de Sitter regions, since as $r \rightarrow
\infty$ the cosmological term dominates. In other words, for very
large values of the radial coordinate $r$ the large-scale curvature
of the spacetime must be taken into account~\cite{LemosLobo}. On the
other hand, it is remarkable that in this case we can have positive
energy density not only for $\omega_r<0$, and also for positive
values of the state parameter $\omega_r$ by requiring that $\Lambda
> - \frac{(N-1)(2-N)}{4 \omega_r r_0^{2}} (N(\omega_r+1)-2)$.

It is interesting to note that if we require that the lateral
pressure has also a barotropic equation of state given by
Eq.~(\ref{12A}), i.e. $p_l(r)=\omega_l \rho(r)$, then we obtain that
the cosmological constant must vanish, i.e. $\Lambda=0$, and the
dimensional constraint
\begin{equation}
N(\omega_l+1)-\omega_l + \omega_{{r}}=2
\end{equation}
must be fulfilled. In this case the lateral pressure may be written
as
\begin{eqnarray}
p_l(r)=\left(\frac{2-N-\omega_{{r}}}{N-1} \right) \, \rho(r)=
\nonumber \\ \frac{\left(2-N-\omega_r \right) \left(2- N \right)}{2
\omega_r r_0^{2}} \,{\left(\frac{r_0}{r}\right)}^{{\frac
{N(\omega_{{r}}+1)-2}{\omega_{{r}}}}}.
\end{eqnarray}
Thus, static wormhole configurations with $e^{\phi(r)}=1$ and
anisotropic barotropic pressure with constant state parameters are
always, in any dimension $N \neq 3$, asymptotically flat spacetimes.

\subsection{Pressure with constant state parameters}
Now we shall consider the case $\phi(r) \neq 0$ and barotropic
pressure with constant state parameters~(\ref{12}) and~(\ref{12A}).
This means that now we have three field equations for three unknown
functions $b(r)$, $\phi(r)$ and $\rho(r)$.

By putting $p_l(r)=\omega_l \rho(r)$ into Eq.~(\ref{staticconseq})
we obtain that the energy density is given by
\begin{eqnarray}\label{exppp}
\rho(r) = C \, r^{\frac{(N-1)(\omega_r-\omega_l)}{\omega_r} }  \,
{e^{-{\frac { \left( 1+\omega_{{r}} \right) \phi \left( r \right)
}{\omega_{{r}}}}}},
\end{eqnarray}
where $C$ is an integration constant. From Eq.~(\ref{static9}), and
by taking into account the Eq.~(\ref{exppp}), we find that
\begin{eqnarray}\label{b(r)}
b(r)= 2{r}^{-N+3}\int {\frac {\left( \kappa \, C \,
\frac{r^{\frac{(N-1)(\omega_r-\omega_l)}{\omega_r}}}{ {e^{{\frac {
\left( 1+\omega_{{r}} \right) \phi \left( r \right)
}{\omega_{{r}}}}}} } +\Lambda \right) {r}^{N-1}}{N-1}}{dr}+
\nonumber
\\ { C_1} {r}^{-N+3}, \,\,\,\,\,\,\,\,\,\,\,\,\,\,\,\,
\end{eqnarray}
where $C_1$ is a new integration constant. In order to have the
general solution for this case we can find the function $\phi(r)$ by
solving the differential equation~(\ref{static3}), by putting into
it the expressions~(\ref{exppp}) and~(\ref{b(r)}). Unfortunately,
the obtained differential equation for $\phi(r)$ is too complicated,
so we shall give a particular solution by considering a restricted
choice of the shift function $\phi(r)$.

Let us consider the case
\begin{equation}\label{shift f}
e^{\phi(r)}=\left(\frac{r}{r_0} \right)^\alpha.
\end{equation}
This choice clearly may ensure the absence of horizons and
singularities for $0<r_0 \leq r < \infty$, since the shift function
is finite throughout the space-time. In this case the field
equations require that $\Lambda=0$, so in order to have a shift
function of the form~(\ref{shift f}) the cosmological constant must
vanish.
Thus the metric takes the form
\begin{eqnarray}\label{wwormhole}
ds^2=-\left(\frac{r_0}{r} \right)^{N-2}dt^2+\frac{dr^2}{1+{\frac
{1}{\omega_{{r}}}}-\frac{C}{\left({\frac{r}{r_0}}\right)^{N-2}}}+
\nonumber
\\ r^2 d\Omega^2_{N-1},
\end{eqnarray}
where $C$ is a constant of integration, the energy density is given
by
\begin{eqnarray}
\kappa \rho(r)&=&-{\frac {\left( N-1 \right)  \left( N-2 \right) }{2
r_0^2 \omega_{{r}}\left(\frac{r}{r_0} \right)^2}},
\end{eqnarray}
and the constraint
\begin{eqnarray}\label{constraint}
\omega_{{l}}=\frac {-N+N\omega_{{r}}+2-4\,\omega_{{r}}}{2(N-1)}
\end{eqnarray}
was used. Clearly for dimensions $N \geq 3$ we have a positive
energy density if $\omega_r <0$, and a negative energy density for
$\omega_r>0$.

One can rewrite the wormhole metric~(\ref{wwormhole}) in a more
appropriate form. From the condition $g^{-1}_{rr} (r = r_0) = 0$, we
obtain for the integration constant $C=1+1/\omega_r$. Then, we have
for the metric component
$g_{rr}^{-1}(r)=(1+1/\omega_r)(1-1/(r/r_0)^{(N-2)})$, and the
wormhole throat is located at $r_0$.

Note that this wormhole is not asymptotically flat, so in order to
this wormhole connects two different asymptotically flat regions we
need to match this solution to an exterior N-dimensional vacuum
spacetime, i.e. to the N+1-dimensional Schwarzschild
solution~\cite{Myers}.

This wormhole spacetime has an interesting feature to be remarked:
if we rescale the coordinate time $t$ and the radial coordinate $r$
we can rewrite the metric to
\begin{eqnarray}\label{deficit wormhole}
ds^2=-\left(\frac{\tilde{r}_0}{r}
\right)^{N-2}dt^2+\frac{dr^2}{1-\frac{1}{\left({\frac{r}{\tilde{r}_0}}\right)^{N-2}}}+
\nonumber
\\  \left(1+\frac{1}{\omega_r} \right) r^2 d\Omega_{N-1}^2,
\end{eqnarray}
This metric has a solid angle deficit, which depends on the value of
the state parameter.

It is interesting to note that if we want an isotropic solution with
the shift function of the form~(\ref{shift f}) we obtain from
Eq.~(\ref{constraint}), by putting $\omega_r=\omega_l$, that
\begin{eqnarray}\label{omegaN}
\omega=\frac {2-N}{2+N}.
\end{eqnarray}
Thus the metric is given by
\begin{eqnarray}\label{n dim sol1}
ds^2=-\left(\frac{r_0}{r} \right)^{N-2}dt^2+\frac{dr^2}{{\frac
{4}{2-N}}-\frac{C}{\left({\frac{r}{r_0}}\right)^{N-2}}}+  r^2
d\Omega_{N-1}^2, \nonumber \\
\end{eqnarray}
where the energy density and the isotropic pressure are given
\begin{eqnarray}
\rho(r)&=& {\frac {\left( N-1 \right)  \left( N+2
\right) }{2 \kappa r_0^2 \left(\frac{r}{r_0} \right)^2}}, \\
p(r)&=& {\frac {\left( N-1 \right)  \left( 2-N \right) }{2 \kappa
r_0^2 \left(\frac{r}{r_0} \right)^2}}.\label{n dim sol2}
\end{eqnarray}
This energy density is always positive. In order to have a wormhole
we must require $C=\frac{4}{2-N}$, ensuring that
$g_{rr}^{-1}(r=r_0)=0$. However, it can be shown that in this case
the metric component $g_{rr}$ is positive for $0< r_0< r_0$ and
negative for $r > r_0$, so the metric~(\ref{n dim sol1}) does not
represent a wormhole spacetime for $N \geq 3$.

Let us note that a four dimensional spherically symmetric static
wormhole solution with a shift function of the form~(\ref{shift f})
and isotropic pressure was considered by Lobo in Ref.~\cite{Lobo}.
However the solution given by Eq. (32) of the Ref.~\cite{Lobo} does
not have isotropic pressure of the form $p=\omega \rho$. As we can
see from the N-dimensional solution~(\ref{n dim sol1})-(\ref{n dim
sol2}) the self-consistent four-dimensional solution with a shift
function of the form $e^{\phi}=\left(\frac{r}{r_0}\right)^{\alpha}$
is given by
\begin{eqnarray}\label{5 dim sol1}
ds^2=-\left(\frac{r_0}{r} \right)
dt^2+\frac{dr^2}{{-4}-\frac{C}{{\frac{r}{r_0}}}}+ r^2 d \Omega_2^2,
\end{eqnarray}
where the isotropic pressure and the energy density are given
$p=-\frac{1}{5}  \rho(r)=-{\frac {1}{\kappa r_0^2
\left(\frac{r}{r_0} \right)^2}}$l, and as we stated above this
solution is not a wormhole.

\section{Evolving $N+1$ Wormhole solutions}
Now we shall consider the non-static branch of the solutions. As we
stated above in order to have non-static wormholes we must require
$\partial \phi(t,r)/\partial r=0$. This condition implies that the
redshift function can only be a function of t, i.e. $\phi(t, r) =
f(t)$, and then we can rescale the time coordinate $t$, so without
any loss of generality we shall require $\phi(t,r)=f(t)=0$. In this
case we are interested in solutions having a barotropic anisotropic
pressure with constant state parameters given by Eqs.~(\ref{12})
and~(\ref{12A}).

Now, from the conservation equation~(\ref{ConsEQ2}) we obtain
\begin{equation}\label{5}
\rho(t,r)=\rho_0 a^{-\left(\omega_r +(N-1)\omega_l
+N\right)}r^{\frac{N-1}{\omega_r}\left(\omega_l -\omega_r\right)},
\end{equation}
where $\rho_0$ is an integration constant, and by subtracting
equations (\ref{3}) and (\ref{4}) we have that
\begin{equation}\label{EVWbder}
\frac{N-2}{2a^2 r^3}\left(rb'-3b\right)=\frac{\kappa \rho_0
\left(\omega_r -\omega_l\right)r^{\frac{N-1}{\omega_r}\left(\omega_l
-\omega_r\right)}}{a^{\left(\omega_r +(N-1)\omega_l +N\right)}}.
\end{equation}
\paragraph*{}It is straightforward to see that in order to have a solution
for the shape function $b = b(r)$ the following constraint must be
imposed;
\begin{equation}\label{6}
\omega_r +(N-1)\omega_l +N=2,
\end{equation}
on the state parameters $\omega_r$ and $\omega_l$, thus obtaining
for the shape function and the energy density
\begin{eqnarray}
b(r)&=&C_1 r^3-\frac{2\kappa \rho_0 \omega_r}{(N-2)(N-1)} \,
r^{-\frac{N-2}{\omega_r}-(N-3)}, \,\,\,\, \label{7}
\\ \nonumber \\ \rho(t,r)&=&\rho_0
\frac{r^{-\frac{N-2+N\omega_r}{\omega_r}}}{a^{2}}, \label{8}
\end{eqnarray}
where $C_1$ is an integration constant. Now, from equation (\ref{9})
we obtain the following differential equation for the scale factor
\begin{equation}\label{10}
\dot{a}=\pm\sqrt{\frac{2\Lambda a^2}{N(N-1)}-C_1}.
\end{equation}
The solution for this equation depends on the signs of the
cosmological constant $\Lambda$ and the integration constant $C_1$,
as we display in Table \ref{tabla1}.
\newline
\begin{table}[h!]
\begin{center}
\begin{tabular}{|c|c|c|}
\hline
$a(t)$ & $C_1$ & $\Lambda$ \\
\hline
$a_0 e^{\pm\sqrt{\frac{2\Lambda}{N(N-1)}}t}$ & $=0$ & $>0$ \\
\hline
$\sqrt{\frac{C_1 N(N-1)}{2\Lambda}}\sin\left(\sqrt{\frac{-2\Lambda}{N(N-1)}}t+\phi_0\right)$ & $<0$ & $<0$ \\
\hline
$\sqrt{\frac{-C_1 N(N-1)}{2\Lambda}}\sinh\left(\sqrt{\frac{2\Lambda}{N(N-1)}}t+\phi_0\right)$ & $<0$ & $>0$\\
\hline
$\sqrt{\frac{C_1 N(N-1)}{2\Lambda}}\cosh\left(\sqrt{\frac{2\Lambda}{N(N-1)}}t+\phi_0\right)$ & $>0$ & $>0$\\
\hline
\end{tabular}
\caption{\label{tabla1}The table shows the possible scale factors
derived from Eq. (\ref{10}).}
\end{center}
\end{table}
\newline
Now, it must be noted that the radial coodinate in this solution may
be rescaled in order to absorb the integration constant $C_1$. In
this case the metric is given by
\begin{eqnarray}
ds^2=-dt^2 + a^2(t)\times \nonumber \\
\Bigg(\frac{dr^2}{1-kr^2+\frac{2\kappa \rho_0
\omega_r}{(N-1)(N-2)}r^{-\frac{(N-2)(1+\omega_r)}{\omega_r}}}+r^2
d\Omega^2_{N-1}\Bigg), \,\, \label{11}
\end{eqnarray}
where $k=0$ for $C_1=0$, $k=-1$ for $C_1<0$ and $k=1$ for $C_1>0$.
We summarize all possible scale factors for the found wormhole
solutions in Table \ref{tabla2}.
\begin{table}[h!]
\begin{center}
\begin{tabular}{|c|c|c|}
\hline
$a(t)$ & $k$ & $\Lambda$ \\
\hline
$const$ & $0$ & $0$ \\
\hline
$t+const$ & $-1$ & $0$ \\
\hline
$a_0 e^{\pm\sqrt{\frac{2\Lambda}{N(N-1)}}t}$ & $0$ & $>0$ \\
\hline
$\sqrt{\frac{-N(N-1)}{2\Lambda}}\sin\left(\sqrt{\frac{-2\Lambda}{N(N-1)}}t+\phi_0\right)$ & $-1$ & $<0$ \\
\hline
$\sqrt{\frac{N(N-1)}{2\Lambda}}\sinh\left(\sqrt{\frac{2\Lambda}{N(N-1)}}t+\phi_0\right)$ & $-1$ & $>0$\\
\hline
$\sqrt{\frac{N(N-1)}{2\Lambda}}\cosh\left(\sqrt{\frac{2\Lambda}{N(N-1)}}t+\phi_0\right)$ & $1$ & $>0$\\
\hline
\end{tabular}
\caption{\label{tabla2}The table shows all the possible scale
factors for the general solution (\ref{11}) of an evolving
Lorentzian wormhole in $N+1$ dimensions with the radial tension and
the tangential pressure having barotropic equations of state with
constant state parameters.}
\end{center}
\end{table}
\newline
Now we shall rewrite the wormhole metric (\ref{11}) in a more
appropiate form. From the condition $g_{rr}^{-1}(r=r_0)=0$, we
obtain for the integration constant $\rho_0$
\begin{equation}
\rho_0=\frac{(N-1)(N-2)(kr^2_0-1)}{2\kappa
\omega_r}r_0^{\frac{(N-2)(1+\omega_r)}{\omega_r}},
\end{equation}
yielding for the shape function and the metric component $g_{rr}$
\begin{eqnarray}
b(r)&=&r_0\left(\frac{r}{r_0}\right)^{-\frac{N-2+(N-3)\omega_r}{\omega_r}}+ \nonumber \\
&& kr^3\left(1-\left(\frac{r}{r_0}\right)^{-\frac{N(1+\omega_r)-2}{\omega_r}}\right), \\
a^2(t)g_{rr}^{-1}&=&1-\left(\frac{r}{r_0}\right)^{-\frac{(N-2)(1+\omega_r)}{\omega_r}}- \nonumber \\
&&
kr^2\left(1-\left(\frac{r}{r_0}\right)^{-\frac{N(1+\omega_r)-2}{\omega_r}}\right),
\label{53A}
\end{eqnarray}
respectively. This implies that the wormhole throat is located at
$r_0$, and the energy density is given by
\begin{equation}
\kappa \rho(t,r)=\frac{(N-1)(N-2)(kr^2_0-1)}{2 \omega_r r_0^2
a^2(t)} \left(\frac{r_0}{r}
\right)^{\frac{N(1+\omega_r)-2}{\omega_r}}.
\end{equation}
It is clear that for $N \geq 3$ the cosmological constant directly
controls the behavior of the scale factor $a(t)$ and not the shape
function $b(r)$, which  mainly is controlled by the state parameter
$\omega_r$. In order to have an evolving wormhole we must require
$\omega_r < -1$ or $\omega_r > 0$ (in both of these cases, in the
$g_{rr}$ metric component~(\ref{53A}), $(N-2)(1+\omega_r)/\omega_r
> 0$ and $(N(1+\omega_r)-2)/\omega_r> 0$), implying that the phantom
energy can support the existence of evolving wormholes in the
presence of a cosmological constant, and the energy density vanishes
at $r \rightarrow \infty$. On the other hand, clearly for
$\omega_r<-1$ or $\omega_r>0$ the metric~(\ref{11}) at spatial
infinity ($r \rightarrow \infty$) has slices $t=const$ which are
N-dimensional spaces of constant curvature: open for k = -1, flat
for k = 0 and closed for k = 1. This implies that for $r \rightarrow
\infty$ the metric~(\ref{11}) is foliated with spaces of constant
curvature.

Now some words about the energy conditions. It is well known that,
in all cases, the violation of the weak energy condition (WEC)
\begin{eqnarray}
\rho \geq0, \,\,\, \rho +p_r \geq0, \nonumber \\
\rho +p_{_l} \geq0,
\end{eqnarray}
is a necessary condition for a static wormhole to exist. In the case
of our non-static solution we must require $\omega_r<-1$ or
$\omega_r>0$ in order to have evolving wormholes. Thus in general,
for $k=0$ or $k=-1$ and $\rho>0$, we must require $\omega_r<-1$, so
the WEC is always violated (this is independent of the value of the
cosmological constant), while for $k=1$ (in this case $\Lambda
>0$) and $\rho>0$, we may require $\omega_r<-1$ for $r^2_0 <1$ (and
the WEC is always violated) or require $0<\omega_r<1$ for $r^2_0 >1$
(and the violation of WEC is avoided). Unfortunately this latter
case must be ruled out for the consideration of evolving wormhole
configurations.

For $N=3$ we obtain the evolving wormhole solutions discussed in
Refs.~\cite{Cataldo1} and~\cite{Cataldo2}, where the traversability
criteria for these four dimensional wormholes are also considered.

\section{$2+1$ evolving Lorentzian wormholes}
As we can see from evolving N+1 wormhole solutions the shape
function $b(r)$ in Eq.~(\ref{7}) is not well defined for $N=2$. On
the other hand, by studying more accurately the field equations for
evolving wormholes in any dimensions~(\ref{9})-(\ref{4}) we conclude
that the nature of such wormholes for $N=2$ and $N \geq 3$, are
quite different. Effectively, in this case the condition $\phi(r)=0$
must be required. Thus, the pressure sustaining the tree-dimensional
evolving wormholes must be always isotropic and homogeneous, i.e. of
the form $p_r(t)=p_l(t)=p(t)$, while for $N \geq 3$ the pressure
must be always inhomogeneous and anisotropic, i.e. given by
$p_r(t,r)$ and $p_l(t,r)$. In other words, it can be seen from the
field equations~(\ref{9})-(\ref{4}) that for $N \geq 3$ the
requirements $\phi(r)=0$, $p_r=p_r(t)$ and $p_l=p_l(t)$ immediately
implies that the shape function $b(r)$ must vanish, and then we must
consider a pressure of the form $p_r(t,r)$ and $p_l(t,r)$. For $N=2$
clearly this does not occur.

So now we shall discuss separately wormhole spacetimes in 2+1
dimensions.

In three dimensional gravity the metric for an evolving wormhole is
given by
\begin{equation}
ds^2=-e^{2\phi(t,r)}dt^2+a^2(t)\left(\frac{dr^2}{1-\frac{b(r)}{r}}+r^2d\theta^2\right).
\end{equation}
The field equations may be directly obtained by putting $N=2$ into
Eqs.~(\ref{9})-(\ref{ConsEQ2}). As in the N-dimensional case the
solutions are separated into two branches: one static branch given
by the condition $a(t)=const$ and another non-static branch for
$\partial \phi(t,r)/\partial r=0$.

\subsection{Static three-dimensional branch}
In general, for the static branch the solutions
may be obtained by putting $N=2$ into Eqs.~(\ref{eeb})
and~(\ref{eephi}), and then we can impose a restricted form of the
shape function $b(r)$ or the redsfiht function $e^{\phi(r)}$. For
$\phi(r)=0$ we obtain that the pressure is isotropic and constant,
and in the presence of the cosmological constant takes the value
$\kappa p_r=\kappa p_l=\Lambda$. In this case the energy density is
given by
\begin{eqnarray}
\kappa \rho(r)=\frac{r b^{\prime}-b}{2 r^3}-\Lambda.
\end{eqnarray}
Note that a static wormhole sustained by a pressureless fluid is
only possible in 2+1 (i.e. $\Lambda=0$). For dimensions $N \geq 3$,
the requirements $\phi(r)=0$ and $p_r=p_l=0$ implies that the shape
function must vanish. Let us now consider a specific asymptotically
flat wormhole given by $b(r)=r_0(r/r_0)^{\alpha}$. Thus the
pressureless fluid has an energy density given by
\begin{eqnarray}
\kappa\,\rho(r)=\frac{\left( \alpha-1 \right)}{2 r_0^2
{(r/r_0)}^{3-\alpha}} -\Lambda.
\end{eqnarray}
In this case, we must require $\alpha<1$ in order to have a
three-dimensional wormhole, thus for $\Lambda=0$ the energy density
is always negative. For $\Lambda \neq 0$ we can demand that $\Lambda
< \frac{\alpha-1}{2 r_0^2}<0$ in order to have a positive energy
density for $r \geq r_0$.

On the other hand, for the shape function given by
$b(r)=r_0(r/r_0)^{\alpha}$ and $\phi(r) \neq 0$ the solution has the
form~(\ref{metric230})-(\ref{rho230}) with $N=2$, obtaining a
2+1-dimensional non-traversable wormhole with an event horizon
located at $r_0$. In this case the energy density is always
negative, to the contrary of the real possibility of having a
positive energy density for $N \geq 3$.

It must be noted that three-dimensional static wormhole
configurations are discussed by Perry and Mann in Ref.~\cite{Mann},
where the constraints on the field equations to obtain wormholes are
presented and further constraints on
traversibility are discussed. 

\subsection{Non-static three-dimensional branch}
Let us now discuss the 2+1-non-static branch with $\partial
\phi(t,r)/\partial r=0$. In this case the Einstein field equations
are given by
\begin{equation}\label{14}
H^2+\frac{\left(rb'-b\right)}{2r^3 a^2}=\kappa \rho + \Lambda.
\end{equation}
\begin{equation}\label{15}
-\frac{\ddot{a}}{a}=\kappa p_r-\Lambda.
\end{equation}
\begin{equation}\label{16}
-\frac{\ddot{a}}{a}=\kappa p_l-\Lambda.
\end{equation}
\begin{eqnarray}
\frac{\partial \rho}{\partial t}+H\left(p_r+p_l+2\rho\right)&=&0.\label{17} \\
\frac{\partial p_r}{\partial r}
-\frac{\left(p_l-p_r\right)}{r}&=&0.\label{18}
\end{eqnarray}
It's clear from Eqs.~(\ref{15}) and~(\ref{16}) that only an
isotropic pressure is permitted, then we shall write $p_r=p_l=p$.
Thus from Eq.~(\ref{18}) we obtain that $\partial p/\partial r=0$,
which implies that the pressure has the general form $p=p(t)$. In
the following we shall discuss the cases $p=const$ and $p=p(t)$.

\subsubsection{Case $p=const$ and $\rho(t,r)$} By putting $p=const$
into Eqs.~(\ref{15}) and~(\ref{17}) we obtain for the scale factor
and the energy density
\begin{eqnarray}
a(t)= C_1 \sin(\sqrt{\kappa p -\Lambda } \,\, t)+C_2 \cos(\sqrt{\kappa p -\Lambda} \,\, t), \label{30a}\\
\rho(t,r)= \frac{\rho_{_0}(r)}{a^{2}} -p,
\,\,\,\,\,\,\,\,\,\,\,\,\,\,\,\,\,\,\,\,\,\,\,\,\,\,\,\,\,\,
\label{30}
\end{eqnarray}
respectively. Now by putting the scale factor and energy density
from Eqs.~(\ref{30a}) and~(\ref{30}) into Eq.~(\ref{14}) we obtain
the following expression for the shape function:
\begin{eqnarray}
\frac{b(r)}{r}={r}^{2} \left( {{\it C_1}}^{2 }+{{\it C_2}}^{2}
\right)  \left(\Lambda-\kappa\,p \right)+ \nonumber \\ 2\kappa\,\int
r \rho_{_0} \left( r \right) {dr} +C,
\end{eqnarray}
where $C$ is an integration constant. By giving a restricted form of
$\rho_{_0}(r)$ we can obtain the shape function. One can also impose
a restricted form of $b(r)$ and obtain the form of the function
$\rho_{_0}(r)$ with the help of the expression
\begin{eqnarray}\label{rho00}
\kappa \rho_{_0} \left( r \right)=\frac{1}{2r}\left(\frac{b(r)}{r}
\right)^{\prime}+ \left( {{\it C_1}}^{2 }+{{\it
C_2}}^{2} \right)  \left(\kappa\,p -\Lambda \right), \nonumber \\
\end{eqnarray}
where ${}^{\prime}=d/dr$. Let us consider a specific wormhole
solution given by $b(r)=r_0(r/r_0)^{\alpha}$. Thus we obtain
\begin{eqnarray}
\kappa \rho_{_0} \left( r \right)=\frac{\left( \alpha-1 \right)}{2
r_0^2 {(r/r_0)}^{3-\alpha}}+\left( {{\it C_1}}^{2 }+{{\it C_2}}^{2}
\right) \left(\kappa\,p -\Lambda \right).
\end{eqnarray}
As we know $\alpha<1$ for having $b(r)/r \leq 1$, and in order to
have a positive energy density we must require that $\kappa p >
-\frac{(\alpha-1)}{2 r_0^2 (C_1^2+C_2^2)}+\Lambda$. Note that in
this case even for $\Lambda=0$ we can have $\rho(r)>0$.

It is interesting to note that this wormhole, for every slice
$a(t)=a_0=const$, and for $p=\Lambda=0$ with $\alpha=1/2$ reproduces
the asymptotically flat and static wormhole solution discussed in
Ref.~\cite{Mann}.

\subsubsection{Case $p=p(t)$ and $\rho(t,r)$} By direct integration
of Eq.~(\ref{17}) we have that
\begin{eqnarray}\label{66A}
\kappa \rho(t,r)= \frac{F(r)}{a^2(t)}-\frac{2  \kappa} {a^2(t)} \int
\left(a(t) \, p(t) \, \frac{d a(t)}{dt} \right) dt.
\end{eqnarray}
Since $p(t)$ is an arbitrary function, this implies that the general
form of the energy density is $\rho(t,r)= \frac{F(r)}{a^2(t)}+C(t)$.
By putting the expression $p(t)=-\ddot{a}(t)/a(t)+\Lambda$ into
Eq.~(\ref{66A}) we obtain finally
\begin{eqnarray}\label{66B}
\kappa \rho(t,r)= \frac{F(r)}{a^2(t)}+\left(\frac{\dot{a}}{a}
\right)^2 -\Lambda.
\end{eqnarray}
By substituting $\rho(t,r)$ from the above equation in
Eq.~(\ref{14}) we obtain that
\begin{eqnarray}\label{Frr}
F(r)=\frac{1}{2r}\left( \frac{b(r)}{r}\right)^{\prime}.
\end{eqnarray}
Clearly, for having a solution we must give a restricted form of the
pressure in order to find the form of the scale factor. For example,
for $p(t)=p=const$ we obtain the discussed above
solution~(\ref{30a}), (\ref{30}) and~(\ref{rho00}). If we consider
the pressure given by $p(t)=t^{\alpha}$, then the scale factor takes
the form
$a(t)=C_{{1}}{t}^{1/2(1+\sqrt{1-4\,C})}+C_{{2}}{t}^{1/2(1-\,\sqrt
{1-4\, C})}$ for $\alpha=-2$, and
\begin{eqnarray}
a \left( t \right) ={\it C_1}\,\sqrt {t}{\it BesselJ} \left( \left(
\alpha+2 \right) ^{-1},2\,{\frac {\sqrt
{C}{t}^{1/2\,\alpha+1}}{\alpha +2}} \right) + \nonumber \\ {\it
C_2}\,\sqrt {t}{\it BesselY} \left(  \left( \alpha +2 \right)
^{-1},2\,{\frac {\sqrt {C}{t}^{1/2\,\alpha+1}}{\alpha+2}} \right),
\,\,\,\,\,\,\,\,\,\,\,\,
\end{eqnarray}
for $\alpha \neq -2$, where $BesselJ$ and $BesselY$ are the Bessel
functions of the first and second kinds respectively. In order to
have a traversable wormhole we must give a restricted form of the
shape function satisfying all wormhole constraints, as for example,
$b(r)=r_0 (r/r_0)^{\alpha}$. Clearly in this case from
Eq.~(\ref{Frr}) we have $F(r)<0$ for $\alpha<1$, thus the energy
density~(\ref{66B}) may be negative or positive during the
evolution. This mainly depends on the relation between the terms
$F(r)/a^2$ and $H^2$. As an example let us consider the case of a
power law scale factor $a(t)=t^{\beta}$. In this case the energy
density will be given by
\begin{eqnarray}\label{rho0015}
\kappa \rho(r)=\frac{\left( \alpha-1 \right)}{2 r_0^2
{(r/r_0)}^{3-\alpha} t^{2 \beta}}+\frac{\beta^2}{t^2}  -\Lambda.
\end{eqnarray}
Clearly in this case for $\Lambda \neq 0$ there exist values of the
parameters which ensure the positivity of the energy density during
all evolution of the scale factor. For $\Lambda=0$ we always can
find a value of the cosmic time $t=t_0>0$, where the energy density
vanishes, thus $\rho(t,r)$ may be negative for $0<t <t_0$ or $t>
t_0$.

It must be noticed that the case $p(t)$ and $\rho(t)$ must be
excluded from consideration since in this case we must require that
$b(r)=0$, and then we can not have a wormhole configuration.

\section{Discussion}
In this paper we have obtained N+1-dimensional solutions for the
Einstein field equations which describe static and evolving
spherically symmetric Lorentzian wormholes. In general, for static
wormholes, we require that at least the radial pressure has a
barotropic equation of state of the form $p_r=\omega_r \rho$, where
the state parameter $\omega_r$ is constant, and for evolving
wormholes we also require a barotropic equation of state
$p_l=\omega_l \rho$ with constant state parameter $\omega_l$ for the
lateral pressure.

For static wormholes it is shown that, in any dimension $N \geq 3$,
with $\phi(r)=\Lambda=0$ and anisotropic barotropic pressure with
constant state parameters, they are always asymptotically flat
spacetimes, while in 2+1 gravity the static wormholes may have more
general asymptotic spaces. In this case, the matter sustaining the
three-dimensional wormhole may be only a pressureless fluid.

The nature of evolving wormholes in 2+1-dimensions and
N+1-dimensions, with $N \geq 3$, are quite different. For evolving
wormholes in any dimensions the condition $\phi(r)=0$ must be
required. However, this constraint implies that the pressures
sustaining the tree-dimensional wormhole configurations must be
always homogeneous, i.e. only must depend on the cosmological time,
while for $N \geq 3$ these pressures must be always inhomogeneous,
i.e. of the form $p(t,r)$, in order to have an evolving wormhole
spacetime. This can be seen from the field
equations~(\ref{9})-(\ref{4}) (for $N \geq 3$) by requiring
$\phi(r)=0$, $p_r=p_r(t)$ and $p_l=p_l(t)$. This inmediately implies
that the shape function $b(r)$ must vanish. In the case of evolving
wormholes with $N \geq 3$, the presence of a cosmological constant
leads to an expansion or contraction of the wormhole configurations:
for positive cosmological constant we have wormholes which expand
forever and, for negative cosmological constant we have wormholes
which expand to a maximum value and then recollapse. In the absence
of a cosmological constant the wormhole expands with constant
velocity, i.e without acceleration or deceleration. In 2+1
dimensions the expanding wormholes always have an isotropic and
homogeneous pressure, depending only on the time coordinate.

\section{Acknowledgements}
This work was supported by CONICYT through Grant FONDECYT No.
1080530 (MC), PhD. scholarship No. 21080709 (PM), and by Direcci\'on
de Investigaci\'on de la Universidad del B\'\i o-B\'\i o (MC).


\begin{thebibliography}{2}
\bibitem{Norman}J. Ponce de Leon and N. Cruz, Gen. Rel. Grav. {\bf 32}, 1207 (2000).
\bibitem{Gallo} E.~Gallo, Gen.\ Rel.\ Grav.\  {\bf 36}, 1463-1471 (2004).
\bibitem{Myers} R. C. Myers and M. J. Perry, Ann. Phys. (N.Y.) {\bf 172}, 304
(1986).
\bibitem{Emparan} R. Emparan and H. S. Reall, Phys. Rev. Lett. {\bf 88}, 101101
(2002).
\bibitem{Gonzales–Diaz} P. Gonzales–Diaz, Phys. Lett. B 247, 251 (1990); X. Jianjun and
J. Sicong, Mod. Phys. Lett. 6, 251 (1990).
\bibitem{Cataldo3} M.~Cataldo, P.~Salgado and P.~Minning, Phys.\ Rev.\  D {\bf 66},
124008 (2002).
\bibitem{Bhawal} M.~H.~Dehghani, S.~H.~Hendi, Gen.\
Rel.\ Grav.\  {\bf 41}, 1853-1863 (2009); Bhawal and S. Kar, Phys.
Rev. D 46, 2464 (1992).
\bibitem{Rodrigo} E.~Rodrigo, Phys.\ Rev.\  {\bf D74}, 104025 (2006).
\bibitem{Lemos} G.~A.~S.~Dias and J.~P.~S.~Lemos,
[arXiv:1008.3376 [gr-qc]]; F.~Rahaman, M.~Kalam, S.~Chakraborty,
Gen.\ Rel.\ Grav.\  {\bf 38}, 1687-1695 (2006).
\bibitem{DeBenedictis} A.~DeBenedictis, A.~Das, Nucl.\ Phys.\  {\bf B653}, 279 (2003).
\bibitem{Kar} S. Kar and D. Sahdev, Phys. Rev. D 53, 722 (1996).
\bibitem{Morris1} M.S. Morris, K.S. Thorne and U. Yurtsever, Phys. Rev. Lett. {\bf
61}, 1446 (1988).
\bibitem{Morris} M.S. Morris and K.S. Thorne, Am. J. Phys. {\bf
56}, 395 (1988).
\bibitem{Visser} M. Visser, Lorentzian Wormholes: From Einstein to Hawking, (AIP,
New York, 1995).
\bibitem{Dadhich} N. Dadhich, S. Kar, S. Mukherjee
and M. Visser, Phys. Rev. D {\bf 65}, 064004 (2002).
\bibitem{Lobo} F.~S.~N.~Lobo, Phys.\ Rev.\  D {\bf 71}, 084011
(2005).
\bibitem{LemosLobo} J.~P.~S.~Lemos, F.~S.~N.~Lobo and S.~Quinet de Oliveira,
Phys.\ Rev.\  D {\bf 68}, 064004 (2003).
\bibitem{Cataldo1} M. Cataldo, P. Labra\~na, S. del Campo, J. Cris\'ostomo and P. Salgado, Phys. Rev. {\bf D 78}, 104006 (2008).
\bibitem{Cataldo2} M. Cataldo, S. del Campo, P. Minning and P. Salgado, Phys. Rev. {\bf D 79}, 024005 (2009).
\bibitem{Mann} G.~P.~Perry and R.~B.~Mann, Gen.\ Rel.\ Grav.\  {\bf 24}, 305 (1992).
481 (1988).
\end{thebibliography}
\end{document}